\documentclass[conference]{IEEEtran}
\usepackage{cite}
\ifCLASSINFOpdf
  \usepackage[pdftex]{graphicx}
\else
   \usepackage[dvips]{graphicx}
\fi
\usepackage[cmex10]{amsmath}
\usepackage{amssymb}
\usepackage{array}
\usepackage[caption=false, font=footnotesize]{subfig}
\usepackage{multirow}
\usepackage{threeparttable}
\usepackage{url}
\usepackage{stfloats} % used for putting the double-column equation at the bottom of the page -- Ming
\begin{document}
\title{Enhanced Mobile Digital Video Broadcasting with Distributed Space-Time Coding}

\author{
\IEEEauthorblockN{Ming Liu, Matthieu Crussi\`ere, Maryline H\'elard, Jean-Fran\c{c}ois H\'elard}
\IEEEauthorblockA{Universit\'e Europ\'eenne de Bretagne (UEB)\\
INSA, IETR, UMR 6164, F-35708, Rennes, France \\
Email: \{ming.liu; matthieu.crussiere;  maryline.helard; jean-francois.helard\}@insa-rennes.fr}
\and
\IEEEauthorblockN{Youssef
Nasser}
\IEEEauthorblockA{American University of Beirut\\
ECE Department, Beirut, Lebanon\\
Email: yn10@aub.edu.lb}
}

\maketitle

\begin{abstract}
%\boldmath
This paper investigates the distributed space-time (ST) coding proposals for the future Digital Video Broadcasting--Next Generation Handheld (DVB-NGH) standard. We first theoretically show that the distributed MIMO scheme is the best broadcasting scenario in terms of channel capacity. Consequently we evaluate the performance of several ST coding proposals for DVB-NGH with practical system specifications and channel conditions. Simulation results demonstrate that the 3D code is the best ST coding solution for broadcasting in the distributed MIMO scenario.
\end{abstract}

% \fi
\IEEEpeerreviewmaketitle

\section{Introduction}
In order to meet the ever-increasing demand of mobile digital television (DTV) broadcasting, the Digital Video Broadcasting (DVB) consortium started the standardization process of the Next Generation Handheld specification (DVB-NGH)~\cite{DVB_NGH} at the beginning of 2010.
DVB-NGH will be finalized in the first half of 2012 to acquire the leading position in the future mobile DTV market.

Owing to the future extension frame (FEF) defined in DVB-second generation Terrestrial (DVB-T2)~\cite{DVB_T2_Standard},
DVB-NGH can inherit many state-of-the-art transmission technologies such as low density parity check (LDPC) code, orthogonal frequency-division multiplexing (OFDM)
and, more importantly, can share the hardware as well as the frequency channel in a time division manner with the fixed DTV services.
Being different from DVB-T2, the new DVB-NGH is expected to be able to deliver DTV services to the battery-powered mobile receivers efficiently, flexibly and reliably.
To fulfill these requirements, DVB-NGH incorporates the multiple-input, multiple-output (MIMO) technique aiming at achieving higher throughput and improving the robustness of the mobile reception in severe broadcasting scenarios.

This paper investigates the application of MIMO technique in the DTV broadcasting.
We first show that the distributed MIMO scheme is the best choice among typical broadcasting scenarios from the channel capacity perspective.
With this knowledge, we consequently evaluate several distributed space-time (ST) coding proposals for DVB-NGH.
Simulations with DVB-NGH specifications in realistic channel conditions demonstrate that the 3D code~\cite{Nasser20083D} is the best ST coding scheme.
The research results presented in this paper belong to the framework of the European CELTIC project ``ENGINES''~\cite{ENGINES} which is an active contributor to the standardization of DVB-NGH.

In the sequel, the variables with boldface represent the vectors or matrices;
$\mathbf{A}^{T}$, $\mathbf{A}^{\ast}$ and $\mathbf{A}^{\mathcal{H}}$ denotes the transpose, conjugate and Hermitian transpose of the matrix $\mathbf{A}$;
$a^{\ast}$ is the conjugate of the complex number $a$;
$\mathbb{E}\{\cdot\}$ is the expectation value.%;

\section{Broadcasting Scenarios and Channel Capacities}
\subsection{MIMO-OFDM Transmission Model}
Consider a MIMO transmission with $N_T$ transmit and $N_R$ receive antennas, the channel impulse response of an $L$-tap multipath channel can be written as:
\begin{equation}
  \mathbf{G}=\sum_{l=0}^{L-1}\mathbf{H}_l\ \delta(n-l),
\end{equation}
where
\begin{equation}
\mathbf{H}_l = \left[ {\begin{smallmatrix}
   h_{11}(l) & \cdots  & h_{1N_T}(l)  \\
   \vdots &  \ddots  &  \vdots   \\
   h_{N_R1}(l) & \cdots  & h_{N_RN_T}(l) \\
\end{smallmatrix}} \right]_{N_R\times N_T}
\end{equation}
is an ${N_R\times N_T}$ complex-valued matrix representing the $l$th channel tap of the MIMO channel, where the $(p,q)$th element $h_{p,q}(l)$ is the $l$th tap of the $(p,q)$th channel link from the $q$th transmit antenna to the $p$th receive antenna.

\begin{figure*}[!t]
\centerline{
\subfloat[SISO in SFN]{\includegraphics[width=2.5in]{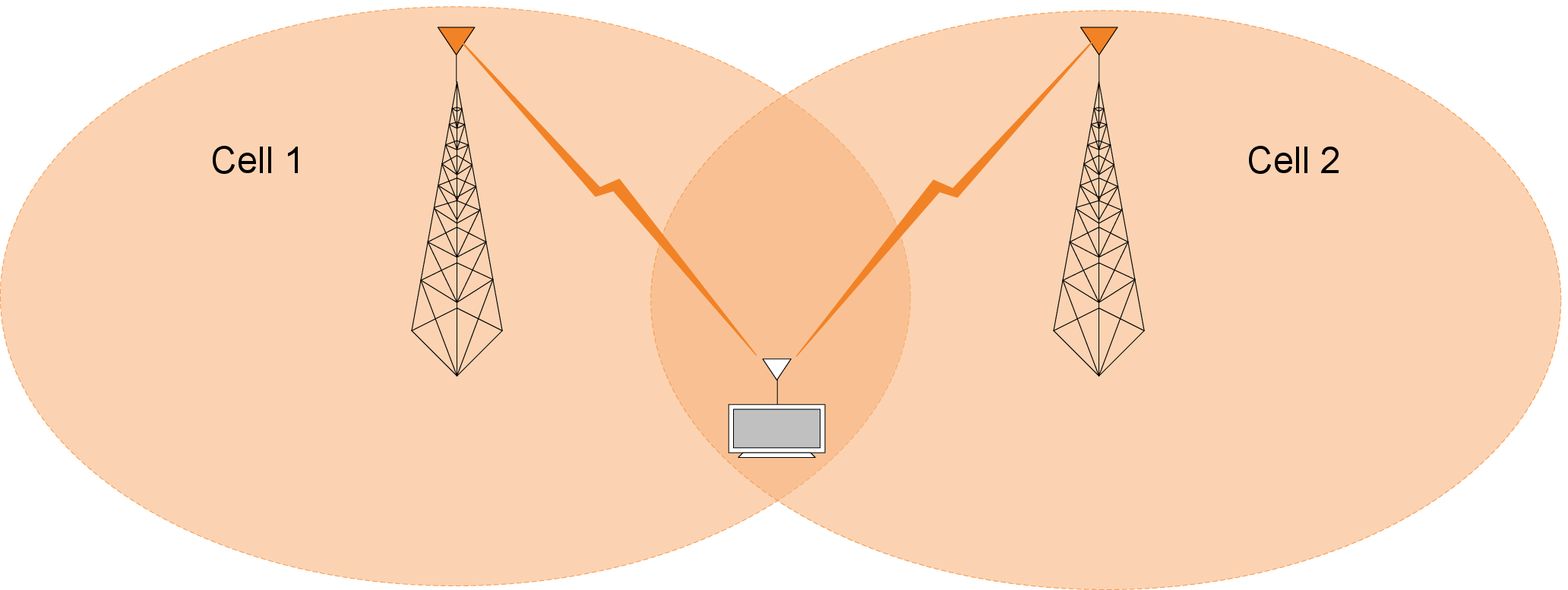}%
\label{fig_SFN}}
\hfill
\subfloat[Sigle cell MIMO]{\includegraphics[width=1.5in]{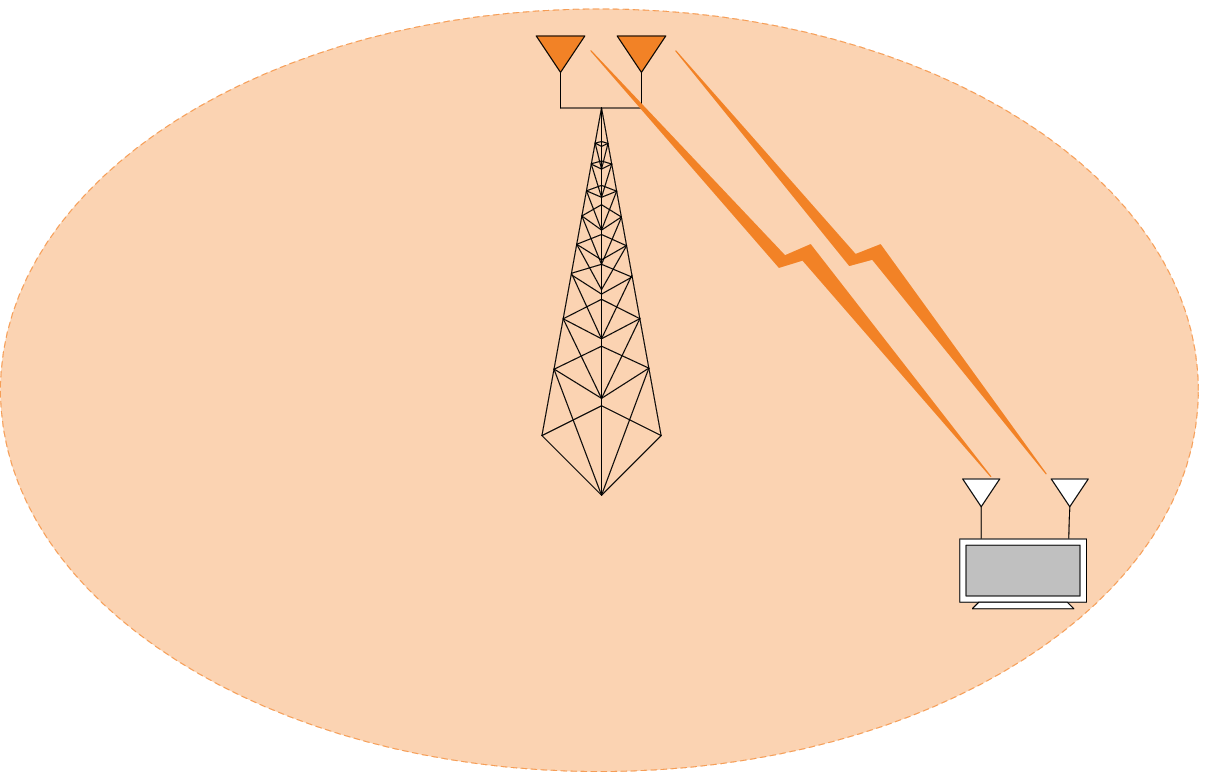}%
\label{fig_MIMO}}
\hfill
\subfloat[Distributed MIMO]{\includegraphics[width=2.5in]{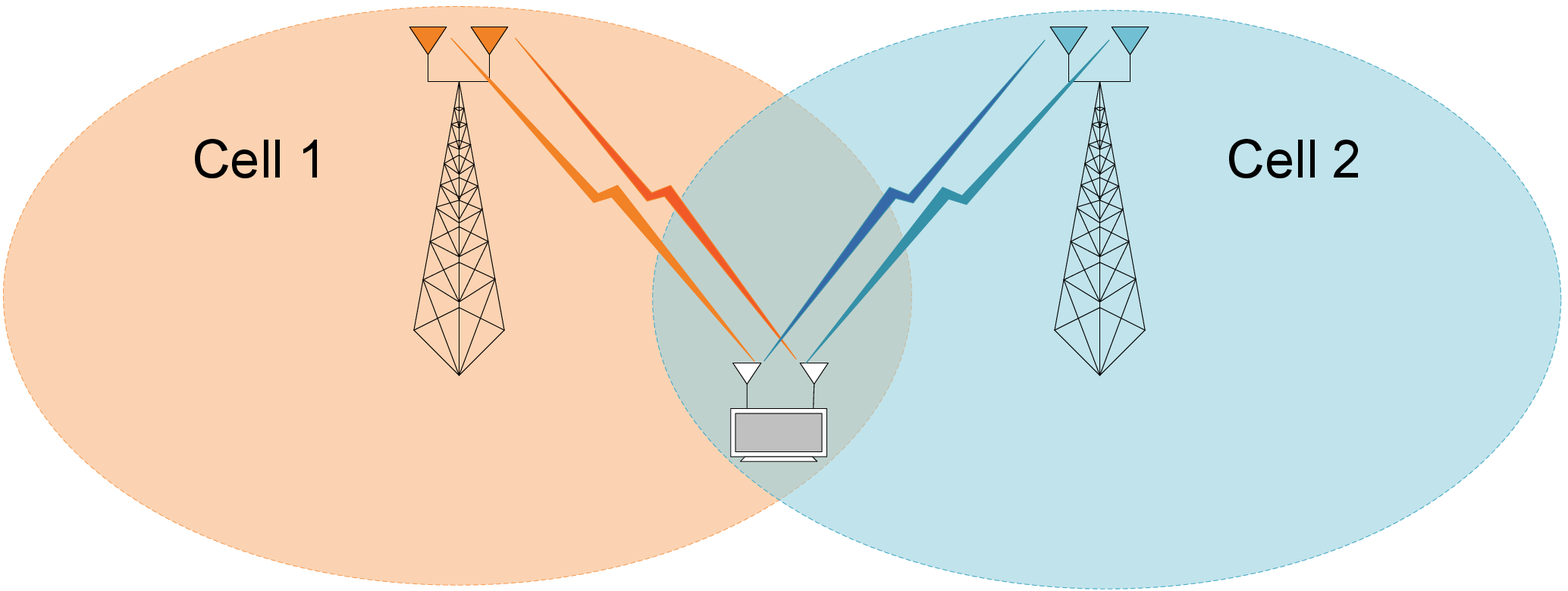}%
\label{fig_distri_MIMO}}
}
\caption{Illustration of different MIMO scenarios.}
%\label{fig_sim}
\end{figure*}

When the cyclic prefix (CP) is long enough compared with the maximum channel delay spread,
the OFDM transmission can be seen as parallel transmissions over a number of flat-fading sub-channels. The channel frequency response for the $k$th subcarrier of the MIMO-OFDM transmission can be written as an $N_R\times N_T$ matrix:
\begin{equation}
\label{eq:Hk}
  \mathbf{H}[k]=\!\!\sum_{l=0}^{L-1}\mathbf{H}_le^{-i\frac{2\pi}{N}kl}\! = \!\!\! \left[ \begin{smallmatrix}
   H_{11}(k) & \cdots  & H_{1N_T}(k)  \\
   \vdots &  \ddots  &  \vdots   \\
   H_{N_R1}(k) & \cdots  & H_{N_RN_T}(k) \\
\end{smallmatrix} \right],
\end{equation}
where the $(p,q)$th element $H_{pq}(k)=\sum_{l=0}^{L-1}h_{p,q}(l)e^{-i\frac{2\pi}{N}kl}$ is the frequency response of $k$th subcarrier through the $(p,q)$th channel link in an $N$-subcarrier OFDM system.
The MIMO-OFDM transmission can be expressed by:
\begin{equation}
\label{eq:rec_sig_mat}
  \mathbf{Y}=\mathbf{H}\mathbf{X}+\mathbf{W},
\end{equation}
where $\mathbf{X}$ is the frequency domain transmitted signal, $\mathbf{Y}$ is the received signal and $\mathbf{W}$ is the additive white Gaussian noise (AWGN). $\mathbf{X}$ is written in stacked vector forms $\mathbf{X} = [X_1(0), \ldots X_{N_T}(0), \ldots X_1(N-1),\ldots X_{N_T}(N-1)]^T$. The same arrangement is applied to $\mathbf{Y}$ and $\mathbf{W}$ as well. The stacked channel matrix is:
\begin{eqnarray}
\label{eq:def_sig_vec}
  && \mathbf{H}= \left[ {\begin{smallmatrix}
   \mathbf{H}[0] & \cdots  & \mathbf{0}  \\
   \vdots &  \ddots  &  \vdots   \\
   \mathbf{0} & \cdots  & \mathbf{H}[N-1] \\
\end{smallmatrix}} \right]_{NN_{R}\times NN_{T}},
\end{eqnarray}
where $X_q(k)$ denotes the signal transmitted on the $q$th antenna and the $k$th subcarrier. Similar notations are applied to $Y_p(k)$'s and $W_p(k)$'s. $\mathbf{W}$ satisfies $\mathbb{E}\{\mathbf{W}\mathbf{W}^{\mathcal{H}}\}=\sigma^2_n\mathbf{I}_{NN_R}$.

\subsection{Broadcasting Scenarios and Capacity Evaluation}

\subsubsection{Single Frequency Network}
The single frequency network (SFN)~\cite{Kateros2009Network} is a spectrally efficient implementation of the broadcasting network.  The same signal is sent from several different transmitters at the same time on the same carrier frequency.
In the following discussion, we focus on the scenario where SFN involves two transmitters
as illustrated by Fig.~\ref{fig_SFN}.
Owing to SFN, the coverage of the broadcasting is expended without the need of additional broadcasting bands.

Considering that the same signal is transmitted from the two transmitters as shown in Fig.~\ref{fig_SFN}, the received signal (\ref{eq:rec_sig_mat}) can be written as:
\begin{equation}
\label{eq:rec_sig_SFN}
 \mathbf{Y} = \underbrace{(\sqrt{\lambda^{(1)}}\mathbf{H}^{(1)}+\sqrt{\lambda^{(2)}}\mathbf{H}^{(2)})}_{\mathbf{H}_{_{\mathrm{SFN}}}}\mathbf{X}+\mathbf{W},
\end{equation}
where $\mathbf{H}^{(1)}$, $\mathbf{H}^{(2)}$, $\mathbf{Y}$, $\mathbf{X}$ and $\mathbf{W}$ follow the definitions in (\ref{eq:def_sig_vec}) with $N_R=N_T=1$ while $\mathbf{H}^{(1)}$ and $\mathbf{H}^{(2)}$ represent the channel matrices associated with the two different transmitters, respectively. $\lambda^{(1)}$ and $\lambda^{(2)}$ are power scale factors of the two channels representing the propagation path losses.

Examining (\ref{eq:rec_sig_SFN}), the SFN transmission can be seen as a SISO transmission with an equivalent channel matrix $\mathbf{H}_{_{\mathrm{SFN}}}$. Keeping the overall transmission power as $P$, the covariance matrix of the transmitted signal is $\boldsymbol\Sigma=(P/2N)\mathbf{I}_{N}$. The ergodic capacity of SFN channel is therefore:
\begin{eqnarray}
 && C_{_{\mathrm{SFN}}} = \mathbb{E}_{\mathbf{H}}\left\{ \! \frac{1}{N}\log_2\Big(\!\det\big(\mathbf{I}_{N} + \frac{1}{\sigma^2_n}{\mathbf{H}_{_{\mathrm{SFN}}}}\boldsymbol\Sigma\mathbf{H}_{_{\mathrm{SFN}}}^{\mathcal{H}} \big)\Big) \right\}  \nonumber\\
 &=&\!\!\!\!\! \mathbb{E}_{\mathbf{H}}\!\!\left\{\!\! \frac{1}{N}\!\! \! \sum_{k=0}^{N-1}\!\log_2\!\!\bigg(\!\!1\!\!+\!\frac{\rho}{2} \Big(\!\lambda^{(1)}\!|H^{(1)}(k)|^2\! \!+ \! \lambda^{(2)}\!|H^{(2)}(k)|^2\!\Big)\! \!\bigg)\! \! \right\},
\end{eqnarray}
where $\rho=P/(N\sigma^2_n)$. The same notations are applied in the following two scenarios.

\subsubsection{MIMO in Single Cell}
Another broadcasting scenario is the implementation of multiple transmit and receive antennas within the same cell. It yields the classical MIMO transmission in the single cell. %MIMO is the only technical solution that can break through the SISO Shannon capacity limit.
Exploring one additional dimension--space domain, MIMO transmission can greatly increase the throughput of the system, namely acquiring \textit{multiplexing gain}.
On the other hand, it can also be used to enhance the reliability of the transmission exploiting~\textit{diversity gain}.
A properly designed MIMO transmission scheme can achieve multiplexing gain or diversity gain or a trade-off between them~\cite{Zheng2003Diversity}.
A simple example of MIMO transmission within a single cell with two transmit and two receive antennas is shown in Fig.~\ref{fig_MIMO}.

In the broadcasting scenario, the channel is unknown at the transmitter but known (by pilot-assisted channel estimation) at the receiver.
Supposing that the transmitted signal $X_q(k)$'s are independent Gaussian variables, for a given the overall transmission power $P$, the mutual information is maximized by transmitting signal with equal power, i.e.  $\boldsymbol\Sigma=\mathbb{E}\{\mathbf{X}\mathbf{X}^{\mathcal{H}}\} = (P/NN_T)\mathbf{I}_{NN_T}$.
Ignoring the spectral efficiency loss due to CP, the ergodic capacity of the MIMO-OFDM channel can be expressed as~\cite{Bolcskei2002capacity}:
\begin{eqnarray}
\label{eq:MIMO_cap}
  &&C = \mathbb{E}_{\mathbf{H}}\left\{ \frac{1}{N}\log_2\Big(\det\big(\mathbf{I}_{NN_R} + \frac{1}{\sigma^2_n}\mathbf{H}\boldsymbol\Sigma\mathbf{H}^{\mathcal{H}} \big)\Big) \right\} \nonumber \\
  &=& \!\!\!\!\! \mathbb{E}_{\mathbf{H}}\!\left\{\!\frac{1}{N}\!\sum_{k=0}^{N-1}\log_2\!\Big(\!\det\big(\mathbf{I}_{N_R} + \frac{\rho}{ N_T}\mathbf{H}[k]\mathbf{H}[k]^{\mathcal{H}} \big)\Big)\right\}.
\end{eqnarray}

\subsubsection{Distributed MIMO}
Besides the ST coding among the antennas of the same transmission sites, the ST coding can also be applied among the antennas of adjacent transmission sites, which yields the distributed MIMO transmission.
In the following discussion, we focus on the distributed MIMO scheme with two transmission sites as illustrated in Fig.~\ref{fig_distri_MIMO}.
We assume that each site is equipped with $N_T$ transmit antennas.
The distributed MIMO channel is composed of two bunches of uncorrelated $N_T\times N_R$ MIMO channels denoted by $\mathbf{H}^{(1)}$ and $\mathbf{H}^{(2)}$.
The received signal can be written as:
\begin{equation}
\mathbf{Y} = \underbrace{[\mathbf{H}^{(1)}\ \mathbf{H}^{(2)}]}_{\mathbf{H}} \underbrace{\left[ {\begin{array}{*{20}c}
   \boldsymbol\Lambda^{(1)}&  0 \\
   0 & \boldsymbol \Lambda^{(2)} \\
\end{array}} \right]}_{\boldsymbol\Lambda}\underbrace{\left[ {\begin{array}{*{20}c}
   \mathbf{X}^{(1)} \\
   \mathbf{X}^{(2)} \\
\end{array}} \right]}_{\mathbf{X}}+\mathbf{W},
\end{equation}
where $\mathbf{X}^{(1)}$ and $\mathbf{X}^{(2)}$ are the signal transmitted from each transmission site, respectively. $\boldsymbol \Lambda^{(j)}=\sqrt{\lambda^{(j)}}\mathbf{I}_{NN_T}$ $(j=1, 2)$ are the power scale matrices representing different propagation path losses associated to the two transmission sites.

The ergodic capacity of the distributed MIMO channel is expressed as:
\begin{eqnarray}
\label{eq:distri_MIMO_cap}
  &&C\! = \mathbb{E}_{\mathbf{H}}\!\!\left\{ \!\! \frac{1}{N}\log_2\Big(\!\det\big(\mathbf{I}_{2NN_R}\! + \! \frac{1}{\sigma^2_n}\mathbf{H}\boldsymbol\Lambda \boldsymbol\Sigma \boldsymbol\Lambda^{\mathcal{H}} \mathbf{H}^{\mathcal{H}} \big)\Big) \right\} \nonumber \\
  &=& \mathbb{E}_{\mathbf{H}}\bigg\{ \frac{1}{N}\log_2\Big(\det\big(\mathbf{I}_{2NN_R} + \frac{P}{2\sigma^2_nNN_T}\nonumber\\
  &\times&\left[ {\begin{array}{*{20}c}
   \lambda^{(1)}\mathbf{H}^{(1)}\mathbf{H}^{(1)\mathcal{H}} &  \mathbf{0}  \\
   \mathbf{0}  & \lambda^{(2)}\mathbf{H}^{(2)}\mathbf{H}^{(2)\mathcal{H}} \\
\end{array}} \right]
   \big)\Big) \bigg\}.
\end{eqnarray}

\subsection{Comparison}
\begin{figure}[!t]
\centering
\includegraphics[width=3in]{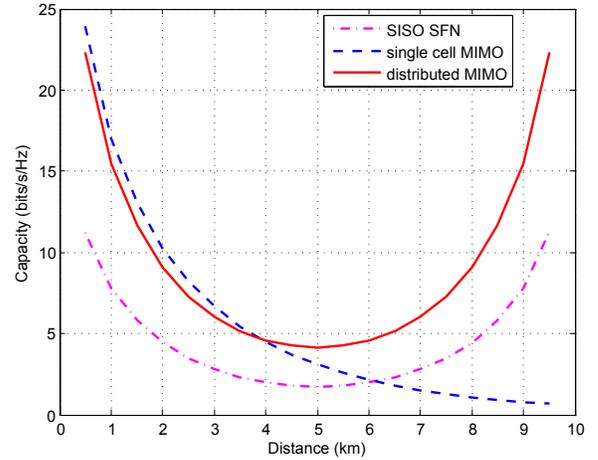}
\caption{Channel capacity comparison among several broadcasting scenarios.}
\label{fig_ch_cap_cmp}
\end{figure}

Fig. \ref{fig_ch_cap_cmp} compares the channel capacities of the three typical broadcasting scenarios discussed above.
This comparison aims at showing the potential of the three broadcasting schemes in terms of the transmission efficiency.
The upper limit of the transmission efficiency is provided for particular broadcasting scheme.
In other words, we evaluate how much throughput can be attained using different transmission schemes with a given amount of transmission power.
Two transmit antennas (one per transmission site) and one receive antenna is considered in the SFN transmission scenario.
Without loss of generality, we choose the number of receive antennas equal to two in the MIMO scenarios.
In the single cell MIMO case, the transmission site equips two transmit antennas. In distributed MIMO case, two transmission sites are involved in our consideration, each having two transmit antennas.
The overall transmission power $P$ is fixed to $10$ kW for all the three transmission schemes.
More precisely, for the single cell MIMO transmission, the transmission power of the cell is $P$ and the power per antenna is $P/2$.
For the SFN and distributed MIMO cases, the transmission power per site is $P/2$.

The channel is assumed to be independent and identically distributed (i.i.d.) Rayleigh channel. That is, all the elements $H_{pq}(k)$'s in (\ref{eq:Hk}) are i.i.d. complex Gaussian random variables with distribution $\mathcal{CN}(0,1)$.
The pathloss model is simply assumed to be:
\begin{equation}
P_r = P\cdot d^{-m},
\end{equation}
where $P_r$ is the received signal power through a propagation distance $d$. It can be seen that the received power decays with $m$th power of the distance $d$. The decaying exponent $m$ is set to $3.5$ which is the typical value of the urban area~\cite{Rappaport2001Wireless}.
The distance between the two transmission sites is assumed to be $10$ km in the SFN and distributed MIMO cases. The two sites locate in the ``$0$ km'' and ``$10$ km'' in Fig. \ref{fig_ch_cap_cmp}, respectively.
In fact, the selection of the distance between transmission sites is related to the network planning~\cite{Kateros2009Network}. Many practical factors should be taken into account, which is beyond the scope of this paper.
Without loss of generality, in our study, the distance is selected so that the SFN can achieve a reasonable minimum capacity, say $1.5$ bits/s/Hz, within its whole coverage.

It can be seen from Fig. \ref{fig_ch_cap_cmp} that the single cell MIMO scheme achieves the highest spectral efficiency in short range.
It is a reasonable results because MIMO technique can acquire multiplexing gain over the classical SISO transmission in high SNR region (i.e. less than $6$ km).
Moreover, since the transmission site of single cell MIMO emits twice transmission power than the distributed MIMO, the single cell MIMO achieves higher capacity than the distributed counterpart in a short range, namely less than $3.5$ km.
Yet, the distributed MIMO scheme obtains a higher average capacity within the whole coverage. Particularly, the distributed MIMO scheme can effectively deliver the high throughput service with a coverage of two SISO broadcasting cells.
In addition, it can significantly improve the capacity at the cell edges (i.e. around $5$ km).
Compared with SISO SFN scheme, the distributed MIMO scheme achieve about twice channel capacity anywhere within the coverage.
In general, \textit{the distributed MIMO scheme is a straightforward extension and effective enhancement of the classical SISO SFN scheme.}

\begin{table*}[t]
\centering
\begin{threeparttable}
\caption{Summary of different ST coding schemes.}
\label{tbl_STs}
\footnotesize
\renewcommand\arraystretch{1.05}
\begin{tabular}{|m{1cm}|m{1.9cm}||m{1cm}|m{0.4cm}|m{0.4cm}|m{1.4cm}|m{1.4cm}|| m{1.65cm}|m{1.65cm}||m{1.5cm}|}
\hline
 \textbf{Category} & \textbf{ST scheme} & \textbf{Nb. of cells} & ${N}_T$ & ${N}_R$ & \textbf{Nb. of info. symb.} & \textbf{Nb. of time solts}& \textbf{Intra-cell ST coding} & \textbf{Inter-cell ST coding} & \textbf{ST decoding complexity} \tnote{a}  \\
\hline\hline
\multirow{4}{1cm}{Classical solutions}& SISO SFN  & 2 & 2 & 1 & 1&1& -- & SFN                  & $\mathcal{O}$($N$) \\ \cline{2-10}
                                        & MISO      & 2 & 2 & 1 & 2&2&-- & Alamouti 2$\times$1& $\mathcal{O}$($N$) \\ \cline{2-10}
                                        & SIMO MRC  & 1 & 1 & 2 & 1&1&-- & --                   & $\mathcal{O}$($N$) \\ \cline{2-10}
                                        & MIMO      & 1 (or 2) & 2 & 2 &2&2& Alamouti (--) & -- (Alamouti) & $\mathcal{O}$($N$) \\
\hline\hline
\multirow{3}{1cm}{{Rate one}}         & Jafarkhani& 2 & 4 & 2 & 4&4&Alamouti& Alamouti        & $\mathcal{O}$($M^4$)\\ \cline{2-10}
                                        & $L_2$ code& 2 & 4 & 2 &4&4& Alamouti-like  & Alamouti   & $\mathcal{O}$($2M^2$)  \\ \cline{2-10}
                                        &Rate 1 Alamouti& 2 & 4 & 2 &2&2& Alamouti      & SFN   & $\mathcal{O}$($M^2$)\\ \hline\hline
\multirow{3}{1cm}{{Rate two}}         & 3D code   & 2 & 4 & 2 &8&4    & Golden    & Alamouti  & $\mathcal{O}$($M^8$)\\ \cline{2-10}
                                        & SM $4\times2$& 2 & 4 & 2 &2&1 & SM $2\times 2$ & SFN  & $\mathcal{O}$($M^2$)  \\ \cline{2-10}
                                        &Rate 2 Alamouti& 2 & 4 & 2 &4&2& Alamouti & SM $2\times 2$ & $\mathcal{O}$($M^4$)\\ \hline
\end{tabular}
{\footnotesize
\begin{tablenotes}
\item [a] The computational complexities required by the rate one and rate two ST coding schemes are the worst case searching times for each received symbol using sphere decoder. The searching space is associated to a given constellation size $M$.
\end{tablenotes}
}
\end{threeparttable}
\end{table*}

\section{ST Coding Schemes with Four Transmit and Tow Receive Antennas}
We investigate six important distributed MIMO coding proposals for the ongoing DVB-NGH standardization~\cite{Jokela2011Performance} in the following sections.

\newcounter{MYtempeqncnt1}
\begin{figure*}[!b]
% ensure that we have normalsize text
\normalsize
% The spacer can be tweaked to stop underfull vboxes.
\vspace*{4pt}
% IEEE uses as a separator
\hrulefill
% Store the current equation number.
\setcounter{MYtempeqncnt1}{\value{equation}}
\setcounter{equation}{14}
\begin{equation}
\label{eq:3D}
\textbf{X}_{\mathrm{3D}}=\left [\begin{array}{*{20}c}
        A & -B^{\ast}\\
        B & A^{\ast} \\
        \end{array}\right]
=\!\frac{1}{\sqrt{5}}\!\left [\begin{smallmatrix}
        \alpha (s_1+\theta s_2) & \alpha (s_3+\theta s_4) & -\alpha^{\ast} (s_5^{\ast}+\theta^{\ast} s_6^{\ast}) & -\alpha^{\ast} (s_7^{\ast}+\theta^{\ast} s_8^{\ast})\\
        i\bar{\alpha} (s_3+\bar{\theta} s_4) & \bar{\alpha} (s_1+\bar{\theta} s_2)  & i\bar{\alpha}^{\ast} (s_7^{\ast}+\bar{\theta}^{\ast} s_8^{\ast}) & -\bar{\alpha}^{\ast} (s_5^{\ast}+\bar{\theta}^{\ast} s_6^{\ast}) \\
        \alpha (s_5+\theta s_6) & \alpha (s_7+\theta s_8) & \alpha^{\ast} (s_1^{\ast}+\theta^{\ast} s_2^{\ast}) & \alpha^{\ast} (s_3^{\ast}+\theta^{\ast} s_4^{\ast})\\
        i\bar{\alpha} (s_7+\bar{\theta} s_8) & \bar{\alpha} (s_5+\bar{\theta} s_6)  & -i\bar{\alpha}^{\ast} (s_3^{\ast}+\bar{\theta}^{\ast} s_4^{\ast}) & \bar{\alpha}^{\ast} (s_1^{\ast}+\bar{\theta}^{\ast} s_2^{\ast}) \\
\end{smallmatrix}
        \right]_{8\times4}
\end{equation}
% Restore the current equation number.
\setcounter{equation}{11}
\end{figure*}

\subsection{Related work}
Since last decades, it has been recognized that higher throughput can be achieved by applying spatial multiplexing~\cite{wolniansky1998v}.
The pioneer work of S. Alamouti~\cite{alamouti1998simple} shows that the orthogonal space-time block code (OSTBC) can extract the spatial diversity with linear processing.
However, full-rate OSTBC only exists for two transmit antennas when complex signal constellations are used.
Various quasi-orthogonal STBCs (QOSTBC) such as~\cite{jafarkhani2001quasi,Hollanti2004Four,belfiore2005golden,Nasser20083D} were proposed by relaxing the requirement of orthogonality.
These QOSTBCs achieve different trade-offs among rate, orthogonality and diversity.
\cite{jafarkhani2001quasi,Hollanti2004Four} proposed group-wise orthogonal codes for four-transmit-antenna cases.
\cite{belfiore2005golden} proposed Golden code, a full diversity, quasi-orthogonal ST code for two-transmit-antenna cases achieving the optimal diversity-multiplexing gain tradeoff.
\cite{Nasser20083D} combined the merits of Alamouti and Golden codes to obtain good performance in distributed MIMO scenarios.
More details and features of these codes are illustrated in the following sections.
The related encoding matrices are given in a hierarchical manner to highlight the schemes for intra-cell and inter-cell ST coding, respectively.

\subsection{Rate one ST codes}

\subsubsection{Jafarkhani code}
A quasi orthogonal ST code is proposed by Jafarkhani in~\cite{jafarkhani2001quasi}. The encoding matrix is:
\begin{equation}\label{eq:Jafarkhani}
\textbf{X}_{\mathrm{Jafarkhani}}\!=\!\!\left [\begin{array}{*{20}c}
        A & -B^{\ast}\\
        B & A^{\ast} \\
        \end{array}\right]\!\!=\!\!\left [\begin{smallmatrix}
        s_1 & -s_2^{\ast} & -s_3^{\ast} & s_4\\
        s_2 & s_1^{\ast}  & -s_4^{\ast} & -s_3\\
        s_3 & -s_4^{\ast} & s_1^{\ast}  & -s_2\\
        s_4 & s_3^{\ast}  & s_2^{\ast}  & s_1 \\
        \end{smallmatrix}\right]_{4\times4},
\end{equation}
where $A$ and $B$ are two successive codewords of Alamouti code~\cite{alamouti1998simple} representing the ST coding carried out among antennas of the same site.
Consequently, $A$ and $B$ are arranged again in an Alamouti manner forming the ST coding among different sites.
The same way of notation is used in the presentation hereafter.

\subsubsection{$L_2$ code}
A similar rate-one code, referred to as $L_2$ code, is proposed in~\cite{Hollanti2004Four}. The encoding matrix is:
\begin{equation}
\label{eq:L2}
\textbf{X}_{L_2}\!\!=\!\!\left [\begin{array}{*{20}c}
        A & -B^{\mathcal{H}}\\
        B & A^{\mathcal{H}} \\
        \end{array}\right]\!\!=\!\!\left [\begin{smallmatrix}
        s_1 & is_2 & -s_3^{\ast} & -s_4^{\ast}\\
        s_2 & s_1  & is_4^{\ast} & -s_3^{\ast} \\
        s_3 & is_4 & s_1^{\ast}  & s_2^{\ast} \\
        s_4 & s_3  & -is_2^{\ast}& s_1^{\ast} \\
        \end{smallmatrix}\right]_{4\times4}.
\end{equation}
Thanks to a modified ``Alamouti-like'' intra-cell coding, the $L_2$ code possesses full-diversity and non-vanishing coding gain.

\subsubsection{Rate one Alamouti code}
Another rate one ST code can be formed by transmitting the same Alamouti codeword in a SFN manner. The encoding matrix can be expressed as:
\begin{equation}\label{eq:R1Alamouti}
\textbf{X}_{\mathrm{R1\ Alamouti}}\!=\!\left [\begin{array}{*{20}c}
        A  \\
        A  \\
        \end{array}\right]\!=\!\left [\begin{smallmatrix}
        s_1 & -s_2^{\ast} \\
        s_2 & s_1^{\ast}  \\
        s_1 & -s_2^{\ast} \\
        s_2 & s_1^{\ast}  \\
        \end{smallmatrix}\right]_{2\times2}.
\end{equation}

\subsection{Rate two ST codes}
\subsubsection{3D code}
A so-called Space-Time-Space (3D) coding is proposed in~\cite{Nasser20083D}.
The intra-cell ST coding is chosen as Golden code, the optimal choice for two-transmit-antenna cases.
The Alamouti scheme is selected as the inter-cell ST coding endowing the overall ST scheme robustness in the presence of transmission power imbalance while preserving the efficiency of Golden code.
The encoding matrix of 3D code is given in (\ref{eq:3D}) (at the bottom of this page), where $\theta=\frac{1+\sqrt{5}}{2}$, $\bar{\theta}=1-\theta$, $\alpha=1+i(1-\theta)$ and $\bar{\alpha}=1+i(1-\bar{\theta})$.

\subsubsection{Spatial Multiplexing}
A simple rate two ST code is formed by transmitting the $2\times2$ spacial multiplexing (SM)~\cite{wolniansky1998v} in a SFN manner:
\setcounter{equation}{15}
\begin{equation}\label{eq:SM}
\textbf{X}_{\mathrm{SM}}\!=\!\left [\begin{array}{*{20}c}
        A  \\
        A  \\
        \end{array}\right]\!=\!\left [\begin{smallmatrix}
        s_1  \\
        s_2  \\
        s_1  \\
        s_2  \\
        \end{smallmatrix}\right]_{2\times1}.
\end{equation}

\subsubsection{Rate two Alamouti code}
Another rate two ST code can be constructed by arranging two independent Alamouti codewords in a SM manner:
\begin{equation}\label{eq:R2Alamouti}
\textbf{X}_{\mathrm{R2\ Alamouti}}\!=\!\left [\begin{array}{*{20}c}
        A  \\
        B  \\
        \end{array}\right]\!=\!\left [\begin{smallmatrix}
        s_1 & -s_2^{\ast} \\
        s_2 & s_1^{\ast}  \\
        s_3 & -s_4^{\ast} \\
        s_4 & s_3^{\ast}  \\
        \end{smallmatrix}\right]_{4\times2}.
\end{equation}

\subsection{Summary of the related ST codes}
The main features of involved ST coding schemes are summarized in Table~\ref{tbl_STs}.
The receiver of the ST coding scheme is more computationally demanding than the classical schemes if the maximum-likelihood (ML) decoding is used.
We should note that the decoding complexity is closely related to the diversity that can be extracted from the ST code depending on different decoding schemes.
For fairness, we consider the complexity that is needed to provide maximum-likelihood (ML) decoding performance.
The decoding of rate two codes is more complex than the rate one counterparts, while information conveyed by the rate two codes is doubled.

\section{Evaluation and Performance Comparison}

\begin{table}[!t]
\renewcommand{\arraystretch}{1.05}
\caption{Simulation Parameters}
\label{tbl_simu_prm}
\centering
\begin{tabular}{|c|c|}
\hline
\textbf{Parameter} & \textbf{Value} \\ \hline\hline
sampling frequency & $9.14$ MHz  \\
FFT size& $4 096$ \\
useful subcarrier& $3 409$ \\
GI length & $1 024$ \\
time interleaver size& $250 K$ cells\\
channel coding& $16 200$-length LDPC, $R=4/9$ \\
LDPC decoding & message-passing algorithm with max 50 iterations \\
\hline
\end{tabular}
\end{table}

\begin{figure}[!t]
\centering
\includegraphics[width=3.5in]{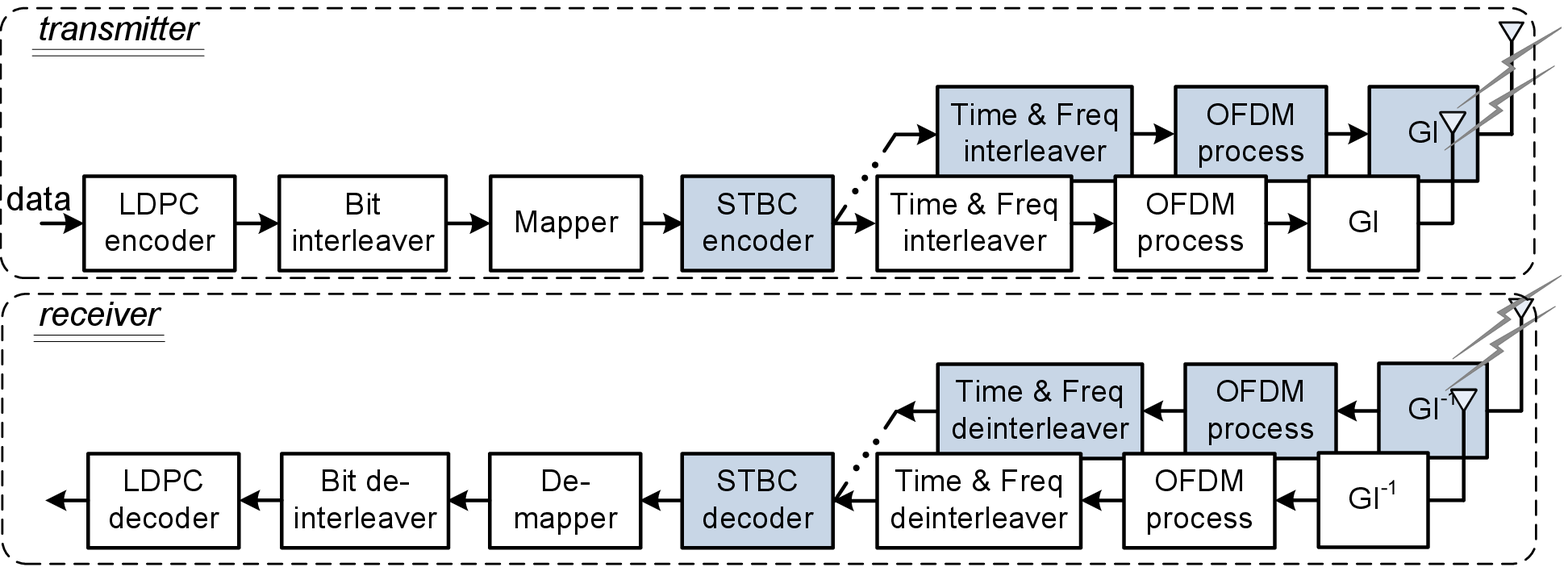}
\caption{Generic block diagram of DVB-NGH. The shaded blocks are the new functionalities of DVB-NGH while others are inherited from DVB-T2.}
\label{fig_block_diag}
\end{figure}

\begin{figure}[!t]
\centering
\includegraphics[width=2.1in]{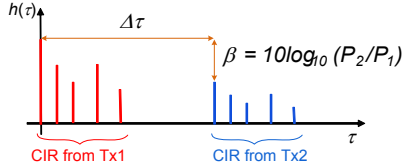}
\caption{Equivalent impulse response of distributed MIMO channel.}
\label{fig_eq_MIMO_CIR}
\end{figure}

In this section we evaluate the performance of different ST coding schemes with the specifications of the DVB-NGH profile.
The block diagram of the DVB-NGH simulation chain is depicted in Fig.~\ref{fig_block_diag}.
Some important simulation parameters are given in Table~\ref{tbl_simu_prm}.
The modulation is selected to be QPSK and 16QAM since the higher order constellations (such as 64QAM and 256QAM) are not the preferred options in the mobile broadcasting scenarios.
The performance of the ST codes is evaluated in both the i.i.d. Rayleigh channel and the novel DVB-NGH  MIMO outdoor channel~\cite{Moss2011NGH}.
The DVB-NGH MIMO channel model emulates the cross-polarized $2\times2$ MIMO transmission in UHF band,
which is realistic and includes many practical transmission and propagation factors including multipath effect, Doppler shift, correlation among channel links etc.
This model also adapts to the distributed MIMO scenario with the combination of two uncorrelated DVB-NGH $2\times2$ MIMO channels.
The channel links related to the farther transmission site is delayed and attenuated by a factor $\beta$ (power attenuation factor) reflecting the effect of the difference of propagation distances as shown in Fig.~\ref{fig_eq_MIMO_CIR}.
The ST decoding algorithm is sphere decoder~\cite{Damen2003maximum} for the distributed MIMO codes.
We assume that the receiver has perfect channel information and is perfectly synchronized.

\begin{figure}[!t]
\centering
\includegraphics[width=3in]{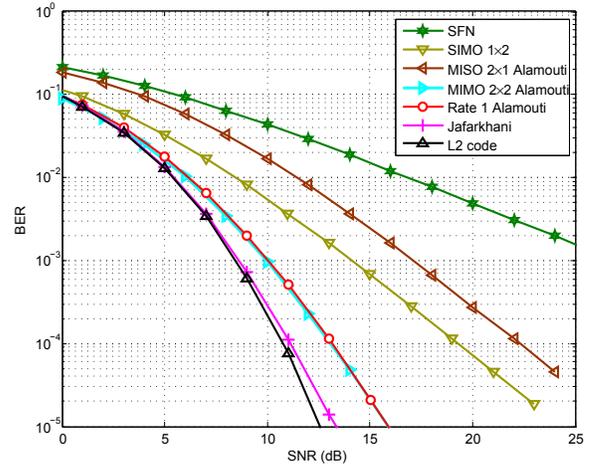}
\caption{Performance of rate one ST codes with QPSK in the i.i.d. Rayleigh channel, no channel coding,
no power imbalance.}
\label{fig_uncoded_BER_Hel_no_imb_r1}
\end{figure}

\begin{figure}[!t]
\centering
\includegraphics[width=3in]{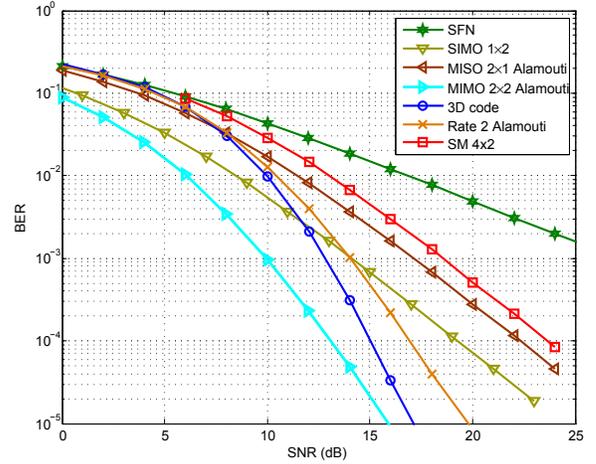}
\caption{Performance of rate two ST codes with QPSK in the i.i.d. Rayleigh channel, no channel coding,
no power imbalance.}
\label{fig_uncoded_BER_Hel_no_imb_r2}
\end{figure}

We first evaluate BER performance of the ST codes without any channel coding in the i.i.d. Rayleigh channel.
The  performances of rate one and rate two codes are given in Fig.~\ref{fig_uncoded_BER_Hel_no_imb_r1} and Fig.~\ref{fig_uncoded_BER_Hel_no_imb_r2}, respectively.
Classical ST coding and diversity schemes are also taken into account as benchmarks in the comparison.
Seen from Fig.~\ref{fig_uncoded_BER_Hel_no_imb_r1}, the distributed MIMO codes with rate one performs better than the classical solutions. This advantage is due to higher diversity obtained by the distributed MIMO codes. %the fact that the distributed MIMO codes achieve higher diversity.
It is reflected by the sharper slop of the BER curves.
The $L_2$ code achieves the best performance among the rate one codes.
Concerning the rate two codes, the 3D code obtains the highest diversity (sharpest BER slop) among all candidates as shown in Fig.~\ref{fig_uncoded_BER_Hel_no_imb_r2}.
We note that the rate two codes obtains twice spectral efficiency as high as the rate one counterparts with the same constellation QPSK.

\begin{figure}[!t]
\centering
\includegraphics[width=3in]{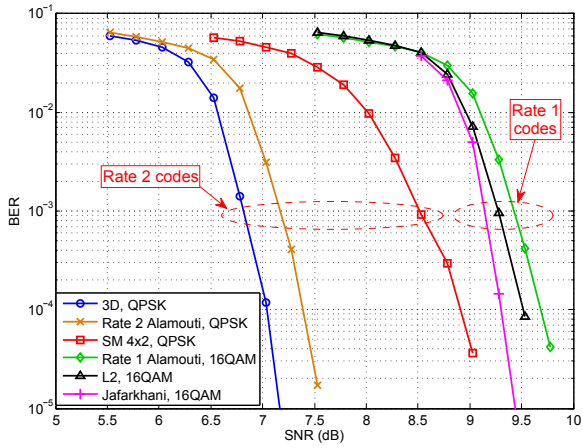}
\caption{Performance of distributed MIMO coding schemes with balanced power, in the DVB-NGH outdoor MIMO channel with $f_d=33.3$Hz.}
\label{fig_BER_Hel_no_imb}
\end{figure}

Consequently, we evaluate the post-LDPC BER performance of ST codes with the same spectral efficient in the DVB-NGH channel.
More precisely, QPSK is used for the rate two codes (3D code, SM and rate two Alamouti) while 16QAM is selected for the rate one codes ($L_2$ code, Jafarkhani code and rate one Alamouti).
It can be observed from Fig.~\ref{fig_BER_Hel_no_imb} that the 3D code outperforms other distributed MIMO coding schemes with the balanced transmission power in the realistic simulation scenario.
It acquires $0.4$ dB and $1.8$ dB gains over rate two Alamouti and SM schemes and more than $2$~dB gains compared to all rate one codes.

\begin{figure}[!t]
\centering
\includegraphics[width=3in]{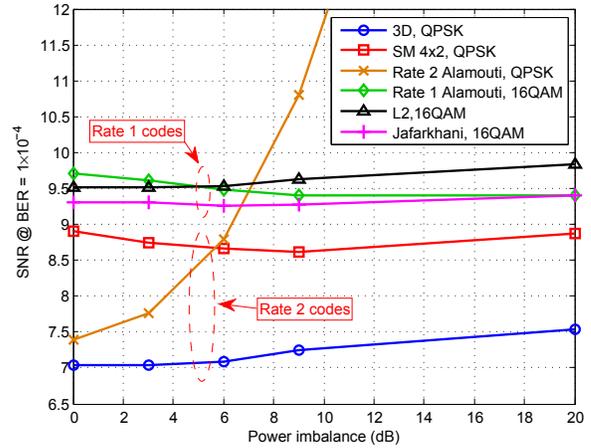}
\caption{Performance of distributed MIMO coding schemes with power imbalance, in the DVB-NGH outdoor MIMO channel with $f_d=33.3$Hz.}
\label{fig_BER_Hel_pw_imb}
\end{figure}

Finally, we investigate the performance of the distributed MIMO codes in the presence of transmission power imbalance.
This study aims at showing the performance of the ST codes in different geographical locations.
Note that we normalize the received signal power to avoid the influence of power loss.
The horizontal axis indicates the ratio of the signal power from the two sites in dB.
It can be seen from Fig.~\ref{fig_BER_Hel_pw_imb} that the rate two Alamouti scheme does not adapt to the power imbalance situation despite its good performance in balanced power case.
This can be explained by the fact that the information delivered by the farther site is totally lost in a strong power imbalance case.
However, the 3D code is the most robust in the presence of power imbalance.
This can be ascribed to the robustness of Alamouti scheme (inter-cell ST coding) in face of strong power imbalance.
In the extreme case ($20$ dB imbalance), the 3D code acquires $1.4$~dB gain over SM scheme and more than $1.9$~dB gains over other rate one ST codes.

\section{Conclusion}
In this paper, we discussed integrating MIMO technique in the digital TV broadcasting, the key topic in the standardization of the DVB-NGH profile. We first analyzed three possible broadcasting scenarios  including SISO SFN, single cell MIMO and distributed MIMO.
We found out that the distributed MIMO is the most promising solution from the prospective of channel capacity.
Consequently, we studied several ST coding schemes that adapt to the distributed MIMO through simulations with the real specifications and the state-of-the-art MIMO channel model of DVB-NGH. Simulation results have shown that the 3D code achieves the best performance among all ST coding schemes in both balanced power and power imbalance cases. The distributed 3D code can be a promising ST coding candidate for the future mobile broadcasting system.

\section*{Acknowledgment}
The authors would like to thank the European CELTIC project ``ENGINES'' for its support of this work.



\bibliographystyle{IEEEtran}
\bibliography{IEEEabrv}

\begin{thebibliography}{10}
\providecommand{\url}[1]{#1}
\csname url@samestyle\endcsname
\providecommand{\newblock}{\relax}
\providecommand{\bibinfo}[2]{#2}
\providecommand{\BIBentrySTDinterwordspacing}{\spaceskip=0pt\relax}
\providecommand{\BIBentryALTinterwordstretchfactor}{4}
\providecommand{\BIBentryALTinterwordspacing}{\spaceskip=\fontdimen2\font plus
\BIBentryALTinterwordstretchfactor\fontdimen3\font minus
  \fontdimen4\font\relax}
\providecommand{\BIBforeignlanguage}[2]{{%
\expandafter\ifx\csname l@#1\endcsname\relax
\typeout{** WARNING: IEEEtran.bst: No hyphenation pattern has been}%
\typeout{** loaded for the language `#1'. Using the pattern for}%
\typeout{** the default language instead.}%
\else
\language=\csname l@#1\endcsname
\fi
#2}}
\providecommand{\BIBdecl}{\relax}
\BIBdecl

\bibitem{DVB_NGH}
``{DVB-NGH, Next Generation Handheld},''
  \url{http://www.dvb.org/technology/dvb-ngh/index.xml}.

\bibitem{DVB_T2_Standard}
{ETSI}, ``{Digital Video Broadcasting (DVB); Frame structure channel coding and
  modulation for a second generation digital terrestrial television
  broadcasting system (DVB-T2)},'' EN 302 755 V1.1.1, Sept. 2009.

\bibitem{Nasser20083D}
Y.~Nasser, J.-F. H\'elard, and M.~Crussi\`ere, ``"{3D MIMO scheme for
  broadcasting future digital TV in single frequency networks},",''
  \emph{Electronics Letters}, vol.~44, no.~13, pp. 829--830, Jun. 2008.

\bibitem{ENGINES}
``{Enabling Next GeneratIon NEtworks for broadcast Services},''
  \url{http://www.celtic-initiative.org/Projects/Celtic-projects/Call7/ENGINES%
/engines-default.asp}.

\bibitem{Kateros2009Network}
D.~A. Kateros, ``{DVB-T Network Planning: A Case Study for Greece},''
  \emph{IEEE Antennas Propag. Mag.}, vol.~51, no.~1, pp. 91--101, Feb. 2009.

\bibitem{Zheng2003Diversity}
L.~Zheng and D.~Tse, ``Diversity and multiplexing: a fundamental tradeoff in
  multiple-antenna channels,'' \emph{IEEE Trans. Inf. Theory}, vol.~49, no.~5,
  pp. 1073--1096, May 2003.

\bibitem{Bolcskei2002capacity}
H.~Bolcskei, D.~Gesbert, and A.~Paulraj, ``{On the capacity of OFDM-based
  spatial multiplexing systems},'' \emph{IEEE Trans. Commun.}, vol.~50, no.~2,
  pp. 225--234, Feb. 2002.

\bibitem{Rappaport2001Wireless}
T.~Rappaport, \emph{{Wireless Communications: Principles and Practice}}.\hskip
  1em plus 0.5em minus 0.4em\relax {Prentice Hall}, 2001.

\bibitem{wolniansky1998v}
P.~Wolniansky, G.~Foschini, G.~Golden, and R.~Valenzuela, ``{V-BLAST: An
  architecture for realizing very high data rates over the rich-scattering
  wireless channel},'' in \emph{Proc. ISSSE}, 1998, pp. 295--300.

\bibitem{alamouti1998simple}
S.~M. Alamouti, ``A simple transmit diversity technique for wireless
  communications,'' \emph{IEEE J. Sel. Areas Commun.}, vol.~16, no.~8, pp.
  1451--1458, Oct. 1998.

\bibitem{jafarkhani2001quasi}
H.~Jafarkhani, ``{A quasi-orthogonal space-time block code},'' \emph{IEEE
  Trans. Commun.}, vol.~49, no.~1, pp. 1--4, Jan. 2001.

\bibitem{Hollanti2004Four}
J.~Hiltunen, C.~Hollanti, and J.~Lahtonen, ``Four antenna space-time lattice
  constellations from division algebras,'' in \emph{Proc. ISIT'04}, 2004.

\bibitem{belfiore2005golden}
J.~Belfiore, G.~Rekaya, and E.~Viterbo, ``The golden code: a 2$\times$2
  full-rate space-time code with nonvanishing determinants,'' \emph{IEEE Trans.
  Inf. Theory}, vol.~51, no.~4, pp. 1432--1436, Apr. 2005.

\bibitem{Damen2003maximum}
M.~Damen, H.~El~Gamal, and G.~Caire, ``On maximum-likelihood detection and the
  search for the closest lattice point,'' \emph{IEEE Trans. Inf. Theory},
  vol.~49, no.~10, pp. 2389--2402, Oct. 2003.

\bibitem{Jokela2011Performance}
T.~Jokela, C.~Hollanti, J.~Lahtonen, R.~Vehkalahti, and J.~Paavola,
  ``{Performance evaluation of 4$\times$2 MIMO schemes for mobile
  broadcasting},'' in \emph{Proc. IEEE BMSB}, june 2011, pp. 1--6.

\bibitem{Moss2011NGH}
P.~Moss, T.~Y. Poon, and J.~Boyer, ``{A simple model of the UHF cross-polar
  terrestrial channel for DVB-NGH},'' \emph{{Research \& Development White
  Paper WHP205}}, Sept. 2011.

\end{thebibliography}
%\begin{thebibliography}{1}
%\end{thebibliography}

\end{document}